\documentclass[conference]{IEEEtran}
\IEEEoverridecommandlockouts

\usepackage{amsmath,amssymb,amsfonts}
\usepackage{graphicx}
\usepackage[full]{textcomp}
\usepackage{xcolor}
\usepackage{multirow}
\usepackage{xurl}
\usepackage{enumitem}
\usepackage{algorithm}
\usepackage{algpseudocode}
\usepackage{bbm}
\usepackage{subcaption}
\usepackage{stix} % circle symbols: https://aneescraftsmanship.com/circle-symbols%E2%97%8B%E2%97%8F%E2%97%8D%E2%97%97%E2%97%94%E2%97%99%E2%A6%BF-in-latex/
\usepackage{amsthm}
\algnewcommand\AND{\textbf{ and }}
\algnewcommand\TO{\textbf{ to }}
\algnewcommand\OR{\textbf{ or }}

\theoremstyle{plain}
\newtheorem{example}{Example}

\theoremstyle{definition}

\renewcommand\IEEEkeywordsname{Keywords}

\begin{document}

\title{A Survey on Coin Selection Algorithms in UTXO-based Blockchains}

\author{\IEEEauthorblockN{Gholamreza Ramezan}
\IEEEauthorblockA{\textit{IEEE Member} \\
g2n@ieee.org}
\and
\IEEEauthorblockN{Manvir Schneider}
\IEEEauthorblockA{\textit{Cardano Foundation} \\
manvir.schneider@cardanofoundation.org}
\and
\IEEEauthorblockN{Mel McCann}
\IEEEauthorblockA{
\textit{Cardano Foundation}\\
mel.mccann@cardanofoundation.org}
}

\maketitle
\begin{abstract}
Coin selection algorithms are a fundamental component of blockchain technology. In this paper, we present a comprehensive review of the existing coin selection algorithms utilized in unspent transaction output (UTXO)-based blockchains. We provide a list of the desired objectives and categorize existing algorithms into three types: primitive, basic, and advanced algorithms. This allows for a structured understanding of their functionalities and limitations. We also evaluate the performance of existing coin selection algorithms. The aim of this paper is to provide system researchers and developers with a concrete view of the current design landscape.
\end{abstract}

\section{Introduction}
\begin{figure}[t!]
\centering
\includegraphics[width=0.45\textwidth]{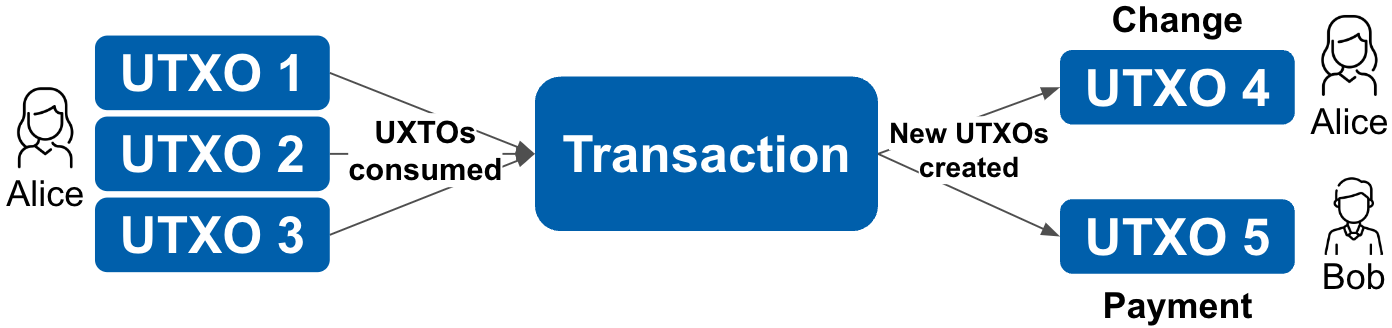}
\caption{Example of new UTXOs created.}
\label{fig_newutxo}
\end{figure}

Coin selection algorithms play a fundamental role in the functioning of blockchains. These algorithms determine which coins or tokens are selected for a particular transaction, effectively shaping the overall efficiency, privacy, and reliability of the blockchain systems. Coin selection is the algorithm of choosing unspent transactions (UTXOs) from a user's wallet in order to pay blockchain tokens to target recipients by forming a transaction. The selected UTXOs set is the input of the transaction while the transaction output includes the payment to the target recipient and the change which goes back to the user's wallet. Fig.~\ref{fig_newutxo} shows a transaction sample with its inputs and outputs. The UTXO model was first introduced in Bitcoin \cite{Nakamoto2008} and Cardano proposed an extended version of UTXO that is called Extended UTXO (EUTXO) \cite{Chakravarty2020}.
%The proper implementation of these algorithms is crucial to maintaining the integrity and usability of blockchain networks, making them an essential component in the evolution and advancement of decentralized technologies

\par Although choosing a set of UTXOs to generate a transaction looks like a simple task, following an effective UTXO selection algorithm is required to avoid long-term challenges. Every time the user selects a set of UTXOs to form inputs for a transaction, the user's wallet may receive some change back as the output of the transaction. This has a smaller amount than the input UTXOs. Over time, if the user continues to pay bills and create transactions, the wallet can end up having too many change UTXOs with small amounts that are called \textit{dust}. A dust UTXO has a small amount that may cost more transaction fees to spend than what it is worth and hence, dust UTXOs are undesired. Dust is analyzed in \cite{dust} and \cite{dust2}.  
\par This paper presents a comprehensive review of the existing coin selection algorithms utilized in UTXO-based blockchains. It is important to note that some established algorithms have not been published formally in the context of coin selection \cite{bitcoin,BitcoinCoinSelectionCode, CoinSelectionForDummies2022PART2, Edsko2018, Erhardt2014, Erhardt2022}. 
Our contributions to this study are threefold. 
\begin{itemize}
    \item We categorize the various existing algorithms based on their characteristics, allowing for a structured understanding of their functionalities and limitations. This provides a valuable resource for researchers and developers seeking to explore and compare different coin selection approaches. 
    \item We provide a comprehensive list of objectives that should be satisfied by coin selection algorithms.
    \item We conduct an in-depth performance analysis and study of existing algorithms, evaluating their effectiveness in terms of factors such as transaction fees, privacy, and overall user experience. The findings of this analysis provide valuable insight into the practical implications and potential benefits of adopting the proposed method in UTXO-based blockchain systems.
\end{itemize}

\par This paper is organized as follows. Section~\ref{sec: Objectives} introduces the terminology and provides a comprehensive list of the objectives required for the coin selection algorithms. Existing algorithms are reviewed in Section~\ref{sec: algos}. The performance of the algorithms is evaluated and the results are provided in Section~\ref{sec: perfomance}. Section~\ref{sec: conclusion} concludes.

\section{Terminology and Objectives} \label{sec: Objectives}
\subsection{Terminology}
\begin{figure}[tbp]
\centering
\includegraphics[width=0.4\textwidth]{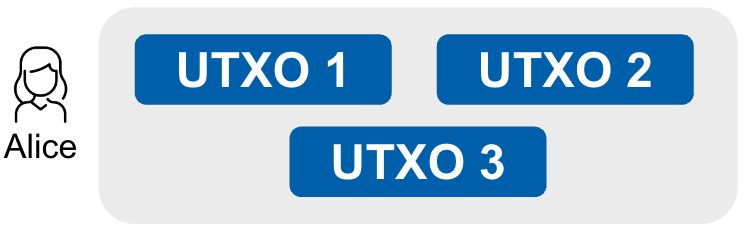}
\caption{User's UTXO pool}
\label{fig_utxopool}
\end{figure}

The inception of the Bitcoin era introduced the need for cryptocurrency wallets. One of the main components of these wallets is the coin selection algorithm, which refers to the selection of a set of UTXOs, denoted $S$, to form an input for a transaction.  Every blockchain user keeps its tokens in the form of UTXOs inside the blocks of a blockchain. A wallet software manages the user's UTXO pool that keeps track of the UTXOs on the blockchain. Fig.~\ref{fig_utxopool} shows the UTXO pool of a blockchain user. Every transaction contains several existing UTXOs as input and new UTXOs as output. The output UTXO has two main parts: the payment UTXO and the change UTXO. The UTXO payment is the token that will be transferred to the receiver's wallet. The value of this UTXO is called the \textit{target value} and is denoted by $T$. The UTXO change is the token that will be sent back to the sender's wallet. The difference between the total values of the input UTXOs and the total values of the output UTXOs is called the \textit{transaction fee}. Fig. \ref{fig_newutxo} shows the payment and the change UTXOs in a transaction. The transaction submitted to a blockchain network will be processed by the block producers. The candidate block producer that picks up a given transaction and includes its UTXOs output into a new block will receive the transaction fee as a part of its block generation reward. 

\par We introduce some notations about UTXOs and UTXO pools. Let $U = \{u_1,...,u_n\}$ be a pool of UTXOs with $n$ UTXOs. For any $i\in [n]$\footnote{Here we use the notation, $[n] = \{1,...,n\}$.}, $u_i^{v}$ is the value, $u_i^{s}$ the size and $u_i^{a}$ the age (confirmation count) of $u_i$. Note that we write $U^{v}$ for the sum of the values of the UTXOs in the set $U$, that is, $U^{v} = \sum_{i \in [n]} u^{v}_i$. Furthermore, we denote $U^{s}$ as the sum of the sizes of the UTXO in the set $U$, that is, $U^{s} = \sum_{i \in [n]} u^{s}_i$. Furthermore, we denote $U^{a}$ as the sum of the UTXO ages in the set $U$, that is, $U^{a} = \sum_{i \in [n]} u^{a}_i$.

\subsection{Objectives}
The coin selection algorithms employ various strategies to ensure efficient utilization of UTXOs while taking into account parameters such as transaction fees, privacy, etc. In this section, we will dive into the objectives that govern the design and implementation of these algorithms. 

\begin{enumerate}[label=(\textbf{O\arabic*}), leftmargin=2.4em]
    \item \label{objective1} Minimizing Transaction Fee: Just pay enough transaction fees to get the transaction included in a block inside a blockchain. A good coin selection algorithm should not only reduce the transaction fee for the current transaction, but should also focus on minimizing the transaction fee in the long run \cite{CoinSelectionForDummies2022}. The lowest number of inputs (UTXOs) is required to minimize transaction size \cite{Wei2023}. The transaction fee may also contain an extra incentive to encourage block producers to accelerate the processing of transactions. 
    \item \label{objective2} Enhancing Privacy: Everyone has access to every transaction. This may result in a breach of financial privacy. Data miners and third-party observers should not be able to get access to user data, such as the total balance or economic activity of a user. A good coin selection algorithm should use a minimum number of UTXO addresses in the selected UTXO set. This is because data miners may use the transaction information to reveal the user's identity \cite{CoinSelectionForDummies2022, Abramova2020}.
    \item \label{objective3} Minimizing Pool Size: The size of the UTXO pool directly affects storage requirements as a way of how storing UTXO affects the processing speed of block producers \cite{Nguyen2018}. In addition, a small UTXO pool limits the number of dust UTXOs in the pool.
    % \item \label{objective4} Increasing Algorithm Performance and Running Speed: Maximizing the speed of execution of the coin selection algorithm and searching for all possible UTXO sets in order to find the best solution \cite{Wei2023}.
    \item \label{objective5} Minimizing Confirmation Time: Block producers select the transaction based on the fee per byte (token per byte) to maximize their revenue. A higher fee will increase the likelihood that a transaction is included in the next block, resulting in faster confirmation \cite{CoinSelectionForDummies2022}.
    \item \label{objective6} Increasing value range of the wallet's UTXO pool: Having a wide value of UTXOs values in a UTXO pool lets the user pay with higher granularity. We explain this in Examples~\ref{example1} and ~\ref{example2}. 
\end{enumerate}
\begin{example}\label{example1}
Assume that there are three users, each has 1 token:
    \begin{itemize}
      \item  User 1 with UTXO pool $U= \{0.5,0.5\}$.\footnote{Note that in numerical examples,  we write $U = \{u_1^v,...,u_n^v\}$ for simplicity.}
      \item  User 2 with UTXO pool $U = \{0.25, 0.25, 0.25, 0.25\}$
      \item User 3 with UTXO pool $U= \{0.1,   0.2,    0.3,   0.4\}$
    \end{itemize}
Now, the set of the possible values of $U$ for each pool is:
    \begin{itemize}
      \item User 1, $\{0.5,1\}$.
      \item User 2, $\{0.25, 0.5, 0.75, 1\}$.
      \item User 3, $\{0.1, 0.2, 0.3, 0.4, 0.5, 0.6, 0.7, 0.8, 0.9, 1\}$.
    \end{itemize}
    As can be seen, User 3 has more options to pay for a wider range of target values. 
    \end{example}
\begin{example}\label{example2} Assume that there are two users, each has 1 token:
    \begin{itemize}
      \item User 4 with UTXO pool $U= \{0.5,0.5\}$.
      \item User 5, with UTXO pool $U= \{0.1, 0.2, 0.3, 0.4\}$.
    \end{itemize}
Now, to pay a target value of $0.6$ token, the coin selection algorithm can form the following UTXO sets as input for the transaction:
    \begin{itemize}
      \item User 4, $\{0.5,0.5\}$, change value $=0.4$.
      \item User 5, $\{0.1,0.2,0.3\}$, change value $=0$.
    \end{itemize}
    As can be seen, for the requested payment from User 4, all of its funds would be used, there would be no confirmed UTXOs left to send another transaction, and the recipient would learn that the sender has an additional $0.4$ token. However, User 5 has more options. A sufficient number of UTXOs in User 5's pool results in not revealing all funds in the wallet. Furthermore, the created transaction would leave the other UTXOs untouched in the wallet and would not require the creation of a change output \cite{Erhardt2022}. 
\end{example}
Another objective that we will not focus on in this work is maximizing the speed of execution of the coin selection algorithm \cite{Wei2023}.

\section{Coin Selection Algorithms}\label{sec: algos}
Various coin selection algorithms are currently employed in wallet and blockchain software, as well as proposed in different papers \cite{Erhardt2016,Nguyen2018,Diroff2019,Abramova2020,Wei2023}. Each coin selection algorithm selects a set of UTXO, denoted by $S^\text{alg} \subseteq U$. The goal is to obtain the optimal value of $S$, which we denote as $S^\text{opt}$. To effectively organize and present the gathered information, we categorize these algorithms into three distinct categories: Primitive, Basic, and Advanced. Each category represents a different level of complexity and sophistication in terms of the strategies used for coin selection. This classification provides a clearer understanding of the evolution and progression of coin selection algorithms, providing valuable information on their effectiveness and potential areas for improvement. In this section, we will dive into each category in detail, exploring the distinctive characteristics and features of the algorithms within them. We summarize the coin selection algorithms in Table~\ref{tab: coin selection algorithms}. 

\begin{table}[t!]
\centering
\caption{Coin Selection Algorithms}
\label{tab: coin selection algorithms}
\begin{tabular}{|l|l|l|}
\hline
\textbf{Category} & \textbf{Coin Selection Algorithm} & \textbf{$S^\text{alg}$} \\
\hline
\multirow{6}{*}{Primitive} & First In First Out (FIFO) & $S^\text{FIFO}$\\
& Last In First Out (LIFO) & $S^\text{LIFO}$\\
& Highest Value First (HVF) & $S^\text{HVF}$\\
& Lowest Value First (LVF)& $S^\text{LVF}$\\
& Highest Priority First (HPF)& $S^\text{HPF}$\\
& Greedy & $S^\text{Greedy}$\\
\hline
\multirow{4}{*}{Basic} & Random Draw \cite{Erhardt2016}& $S^\text{RndDrw}$\\
& Random-Improve \cite{CIP2}& $S^\text{RndImp}$\\
& Knapsack \cite{Erhardt2016}& $S^\text{Knp}$\\
& Branch and Bound \cite{Erhardt2016} & $S^\text{BnB}$\\
\hline
\multirow{4}{*}{Advanced} & Optimization \cite{Nguyen2018} & $S^\text{Opt}$\\
& Knapsack with Leverage \cite{Diroff2019}& $S^\text{KnpLv}$\\
& Myopic and Strategic Optimization \cite{Abramova2020}& $S^\text{Myop}, S^\text{Strat}$\\
& Greedy and Genetic \cite{Wei2023}& $S^\text{GrGe}$\\
\hline
\end{tabular}
\end{table}

\subsection{Primitive Algorithms}
The primitive category refers mainly to algorithms that were discussed and conceptualized during the early stages of the blockchain era. 
% full circle: $\mdlgblkcircle$
% empty circle: $\mdwhtcircle$
% quarter circle: $\circleurquadblack$
% half circle: $\circlerighthalfblack$
% 3/4 circle: $\blackcircleulquadwhite$
%However, as technology evolved, product developers quickly recognized the need for more sophisticated approaches, leading to the emergence of the Basic category.
We start this section by providing a comprehensive comparison of the primitive algorithms. All primitive algorithms mentioned in Table~\ref{tab: coin selection algorithms} are commonly used and known algorithms in computer science. These include First In First Out (FIFO), Last In First Out (LIFO), Highest Value First (HVF), Lowest Value First (LVF), and Highest Priority First (HPF) algorithms. An informal analysis of some of the algorithms is also presented in \cite{Erhardt2014}. Primitive algorithms select a set of UTXOs until the total value of the selected UTXOs reaches the target value. All algorithms in this category use the same approach with different sorting methods. The main approach is depicted in Algorithm~\ref{alg:primitive} where $P(U)$ denotes the alogrithm and $U$ is the sorted UTXO pool. 

\begin{algorithm}
\caption{Primitive Algorithms --- $P(U)$}\label{alg:primitive}
\textbf{Input:} Sorted UTXO pool $U = \{u_1,...,u_n\}$\\
\textbf{Input:} Target $T>0$\\
\textbf{Output:} Set of selected UTXOs $S^\text{alg}$
\vspace{-0.2em}
\begin{algorithmic}[1]
\Require $U^v > T$
    \State $S^{alg} \leftarrow \{\}$
    \State $remain \leftarrow T$
\While{$remain$ $>$ $0$}
\For{$i = 1 \TO n$}
    \State $S^{alg}$.\text{add}($u_i$) 
    \State $remain \leftarrow remain - u_i^v$
\EndFor
\EndWhile
\State \textbf{return} $S^\text{alg}$
\end{algorithmic}
\end{algorithm}

\subsubsection{First In First Out (FIFO)}
The FIFO algorithm picks UTXOs in order of decreasing the confirmation count, that is, older UTXOs are picked first. FIFO has some nice properties, such as the fact that small UTXOs will eventually be spent. The input sets are neither small nor large and, therefore, will not reveal information about the composition of the UTXO pool \ref{objective2}. However, the date of the oldest UTXO in the wallet is revealed. Therefore, it can be guessed how long a user has been using the wallet. The total balance of the wallet might be estimated using the UTXO confirmation count values. It might also be feasible to form a certain pattern using the timestamp ranges of different transactions. Hence, it partially fulfills the second objective \ref{objective2}. FIFO has two additional disadvantages: It minimizes neither the transaction fee \ref{objective1} nor the number of transactions in the UTXO pool \ref{objective3}. We run Algorithm~\ref{alg:primitive} as follows:
\begin{equation*}
    P(\{u_1,...,u_N\}) \ \text{with} \ u_i^a \geq u_{i'}^a \ \text{for} \ i<i'. 
\end{equation*}

\subsubsection{Last In First Out (LIFO)}
The LIFO algorithm selects UTXO in ascending confirmation count order, that is, the newer UTXO are picked first. Similar to FIFO, the input UTXO sets are neither small nor large. Therefore, it will not reveal information about the composition of the UTXO pool. However, there is hardly any consolidation of old UTXOs if the wallet's funds are increasing over time. The objectives \ref{objective1} and \ref{objective3} are not achieved. Furthermore, this algorithm will link up all of the recent activity in the wallet, since the newest change output is always reused. We run Algorithm~\ref{alg:primitive} as follows:
\begin{equation*}
    P(\{u_1,...,u_N\}) \ \text{with} \ u_i^a \leq u_{i'}^a \ \text{for} \ i<i'. 
\end{equation*}

\subsubsection{Highest Value First (HVF)}
The HVF algorithm first picks the UTXO with the highest value. Therefore, only a minimal amount of input will be used, and therefore the transaction fee is minimized \ref{objective1}. HVF will likely increase the size of the UTXO pool, that is, it will not satisfy \ref{objective3}. Furthermore, only a few blockchain addresses are linked. This has a positive impact on privacy \ref{objective2}. However, there remain other privacy issues, as it reveals the upper bound for the UTXO value in the wallet and links consecutive transactions, while the change output is still in the largest UTXO. The algorithm is used in Cardano blockchain during Cardano Improvement Proposal (CIP) 2. We run Algorithm~\ref{alg:primitive} as follows:
\begin{equation*}
    P(\{u_1,...,u_N\}) \ \text{with} \ u_i^v \geq u_{i'}^v \ \text{for} \ i<i'. 
\end{equation*}

\subsubsection{Lowest Value First (LVF)}
The LVF algorithm selects UTXO in ascending value order. Small UTXOs are consolidated as soon as possible, which reduces the wallet's UTXO pool \ref{objective3} and minimizes future spending costs. However, the transaction fee will not be minimized \ref{objective1} since LVF maximizes the input sets. There are privacy concerns as the lower bound for UTXO values in the wallet is revealed. Furthermore, it links consecutive transactions based on the lowest UTXO and also tends to over-consolidate the UTXO pool, which degrades privacy. LVF also links many addresses and decreases the range of values of the UTXO pool of a wallet \ref{objective6}. We run Algorithm~\ref{alg:primitive} as follows:
\begin{equation*}
    P(\{u_1,...,u_N\}) \ \text{with} \ u_i^v \leq u_{i'}^v \ \text{for} \ i<i'.
\end{equation*}

\subsubsection{Highest Priority First (HPF)- Combining FIFO/LIFO with HVF/LVF} The HPF algorithm picks UTXOs in descending order of priority until the target is reached. The priority is calculated as a product of the UTXO values and their age (i.e., the confirmation count). This will decrease transaction fees \ref{objective1} and increase privacy as it links only a few addresses \ref{objective2}. However, this algorithm will not minimize the number of transactions in the UTXO pool \ref{objective3}. We define the priority of $u_i$ as $u_i^p := u_i^v u_i^a$ and run Algorithm~\ref{alg:primitive} as follows:
\begin{equation*}
    P(\{u_1,...,u_N\}) \ \text{with} \ u_i^p \geq u_{i'}^p \ \text{for} \ i<i'. 
\end{equation*}

\subsubsection{A Summary of Primitive Algorithms}
We summarize the primitive methods in Table~\ref{tab: primitive algorithms} and what objectives they achieve and to what extent. As mentioned, in the early days of blockchain development, coin selection algorithms were not extensively researched, and developers often resorted to employing simple algorithms that we reviewed in the primitive category. Although these initial approaches served their purpose to some extent, they lacked the sophistication necessary to handle the complex and evolving demands of blockchain networks. In the next section, we review the basic algorithms.

\begin{table}[t!]
\centering
\caption{ Primitive Algorithms. A white circle $\mdwhtcircle$ indicates that the objective is not fulfilled, while a black circle $\mdlgblkcircle$ is used when the objective is fulfilled. Partial fulfillment is indicated by $\circlerighthalfblack$.}
\label{tab: primitive algorithms}
\begin{tabular}{|l|c|c|c|c|c|}
\hline
\multirow{2}{*}{\textbf{Primitive algorithms}} & \multicolumn{5}{|c|}{\textbf{Objectives}} \\
& \ref{objective1} & \ref{objective2} & \ref{objective3} &  \ref{objective5} & \ref{objective6} \\
\hline
FIFO & $\mdwhtcircle$  & $\circlerighthalfblack$  & $\mdwhtcircle$  &  $ \mdwhtcircle$  & $ \mdwhtcircle$  \\  
LIFO & $\mdwhtcircle$  & $\circlerighthalfblack$  & $\mdwhtcircle$  &  $ \mdwhtcircle$  & $ \mdwhtcircle$  \\  
HVF & $\mdlgblkcircle$  & $\circlerighthalfblack$  & $\mdwhtcircle$  & $ \mdwhtcircle$  & $ \mdwhtcircle$  \\  
LVF & $\mdwhtcircle$  & $\mdwhtcircle$  & $\mdlgblkcircle$  & $ \mdwhtcircle$  & $\mdwhtcircle$  \\ 
HPF& $\mdlgblkcircle$  & $\mdlgblkcircle$  & $\mdwhtcircle$  &  $ \mdwhtcircle$  & $ \mdwhtcircle$  \\
\hline
\end{tabular}
\end{table}

\subsection{Basic Algorithms}

The basic category encompasses coin selection algorithms that replaced their primitive counterparts due to their enhanced efficiency and improved performance. These algorithms represent a significant step forward in terms of usability and effectiveness, meeting the growing demands and challenges of the blockchain field.

\subsubsection{Greedy} 
The greedy algorithm aims to reduce the number of transaction inputs. The algorithm takes UTXOs in descending order and selects UTXOs that are below the remaining target. Once a UTXO is selected, its value is subtracted from the remaining target. The algorithm is described in Algorithm \ref{alg:greedy}.
\begin{algorithm}[t!]
\caption{Greedy Algorithm --- $G(U)$}\label{alg:greedy}
\textbf{Input:} UTXO pool $U = \{u_1,...,u_n\}$ with $u_i^v \geq u_{i'}^v$ for $i<i'$\\
\textbf{Input:} Target $T>0$\\
\textbf{Output:} Set of selected UTXOs $S^\text{greedy}$
\vspace{-0.2em}
\begin{algorithmic}[1]
\Require $U^v > T$
    \State $S^\text{greedy} \leftarrow \{\}$
    \State $remain \leftarrow T $
\For{$i = 1 \TO n$}
\If{$u_i^v \leq$ $remain$ \AND  $remain$ $> 0$}
    \State $S^\text{greedy}$.\text{add}($u_i$) 
    \State $remain \leftarrow remain - u_i^v$
\EndIf
\EndFor
\While{$remain$ $> 0$}
    \State $S^\text{greedy}$.\text{add}($\min\{U \setminus S^\text{greedy}\})$\footnotemark
    \State $remain \leftarrow remain - (\min\{U \setminus S^\text{greedy}\})^v$
\EndWhile
\State \textbf{return} $S^\text{greedy}$
\end{algorithmic}
\end{algorithm}\footnotetext{Note that, whenever we are using a $\min$ or $\max$ function, it is defined as follows: $\min(U) = \{u \in U | u^v \leq \Tilde{u}^v \text{ for all } \Tilde{u}\in U\}$. The $\max$ function is defined analogously.}
Note that the greedy algorithm cannot always minimize the number of inputs. We show this in the following example.

\begin{example}
    Let UTXO pool $U=\{0.25, 0.2, 0.2, 0.1, 0.05\}$ and target $T = 0.4$. The greedy algorithm will find the solution $S^\text{Greedy} = \{0.25, 0.1, 0.05\}$, whereas the optimal solution that minimizes the number of inputs is $S^\text{opt} = \{0.2,0.2\}$. \textcolor{red}{}
\end{example}

\subsubsection{Random Draw}%- Bitcoin Core
The random draw algorithm picks UTXOs randomly with equal probability. The implementation of this method is easy and the method will enhance privacy \ref{objective2} as selected UTXOs have no consistent fingerprint like age or value. Additionally, a random change output leads to an increased value diversity \ref{objective6}. The transaction fee will not be minimized \ref{objective1} and the random draw algorithm can select a UTXO set with an amount greater than the target value. The algorithm is described in Algorithm~\ref{alg:randomdraw} and makes use of a function $Random(U)$ that returns an element of the set $U$ randomly with a uniform distribution.

\begin{algorithm}[t!]
\caption{Random Draw --- $RD(U)$}\label{alg:randomdraw}
\textbf{Input:} UTXO pool $U = \{u_1,...,u_n\}$\\
\textbf{Input:} Target $T>0$\\
\textbf{Output:} Selected UTXOs' set $S^\text{RndDrw}$
\vspace{-0.2em}
\begin{algorithmic}[1]
\Require $U^v > T$
    \State $S^\text{RndDrw} \leftarrow \{\}$
    \State $remain \leftarrow T$
\While{$remain$ $>$ $0$}
    \State $rand \leftarrow (Random(U \setminus S^\text{RndDrw}))^v$
    \State $S^\text{RndDrw}$.\text{add}($rand$) 
    \State $remain \leftarrow remain$ $-$ $rand$
\EndWhile
\State \textbf{return} $S^\text{RndDrw}$
\end{algorithmic}
\end{algorithm}

An alternative implementation for the random draw algorithm is the following. The input UTXO pool $U$ is randomly reshuffled, and the UTXOs are selected starting with the first element of $U$ until the total value of selected UTXOs reaches the target value.

\subsubsection{Random Improve}% - CIP 2} 
The random improve algorithm \cite{CIP2}, as the name suggests, aims at improving the random draw method. To achieve this goal, the algorithm consists of two phases. The first phase is a random draw until there is enough input value to pay for the output. The second phase focuses on improving the selected UTXO set. This is done by expanding the set with additional UTXOs that are chosen one-by-one randomly from the UTXO pool to get as close as possible to twice the target value, $2T$. The reason is as follows. By reaching a value close to twice the target value, we expect a change output with a value that is close to the original target. Unlike tiny change outputs, these ``useful'' outputs will help future transactions to be processed with a lower number of inputs as the previous change outputs are of the value of typical payment requests. The rationale behind phase one is dust management. In particular, if there is a large number of dust UTXOs in the UTXO pool, then with a high probability, a large amount of dust will be selected as input. Therefore, it will reduce the amount of dust in the UTXO pool over time. 

\par Instead of considering only the target $T$, the random improve algorithm considers a target range, consisting of \textit{(low, ideal, high)} $= (T,2T,3T)$. Phase two of the algorithm tries to get as close to the ideal value as possible. The algorithm stops when there is no improvement. There is an improvement after adding a UTXO to the selected inputs if the value of the selected UTXOs is closer to the ideal value than without the additional UTXO. However, it is also necessary that the high value is not exceeded. In addition, there is a maximum input count that has to be considered when adding UTXOs to the selected inputs. The algorithm is described in Algorithm~\ref{alg:randomimprove}. Note that in the algorithm, it is required that the UTXO set is not empty, and it is also allowed to add new UTXOs to the maximum number of input count. The original algorithm in \cite{CIP2} throws an error if the maximum input count is exceeded or the UTXO balance is insufficient. If there are multiple targets to be considered, the original algorithm imposes the constraint that each input can be used only for exactly one target value, otherwise it throws an error. The algorithm will decrease the size of the pool \ref{objective3} and reveal only small information about the wallet, as it uses pseudo-randomly chosen inputs \ref{objective2}. It will not minimize transaction fees \ref{objective1} but could improve the value range of the UTXO pool, as it creates change outputs roughly the size of the target.

\begin{algorithm}[t!]
\caption{Random Improve}\label{alg:randomimprove}
\textbf{Input:} UTXO pool $U = \{u_1,...,u_n\}$\\
\textbf{Input:} Target $T>0$\\
\textbf{Input:} MaxInputCount $MIC$\\
\textbf{Output:} Set of selected UTXOs $S^\text{RndImp}$
\vspace{-0.2em}
\begin{algorithmic}[1]
\Require $U^v > T$
%\textcolor{red}{\Require $|RD(U)|< mic$}
    \State $Selected \leftarrow RD(U)$\footnotemark
    \State $numpPossibleAdditions \leftarrow \min\{MIC, n\} - |Selected|$
\For{$i=1 \TO numPossibleAdditions + 1$}
\If{$i==numPossibleAdditions+1$}
    \State $S^\text{RndImp} \leftarrow {Selected}$
    \State \textbf{break}
\EndIf
    \State $S^\text{RndImp} \leftarrow {Selected}$
    \State $rand \leftarrow Random(U \setminus Selected)$
    \State $Selected$.\text{add}($rand$) 
\If{$|2T - {Selected}^v|>|2T - {S^\text{RndImp}}^v|$}
    \State \textbf{break}
\EndIf
\If{${Selected}^v>3T$}
    \State \textbf{break}
\EndIf
\EndFor
\State \textbf{return} $S^\text{RndImp}$
\end{algorithmic}
\end{algorithm}
\footnotetext{Note that here we assume that $|Selected|< MIC$.}

\subsubsection{Knapsack}% - Bitcoin Core}
The Knapsack algorithm for coin selection was studied by Erhardt~\cite{Erhardt2016}. The algorithm consists of two phases. In the first phase, the algorithm runs through every UTXO in the UTXO pool and adds them one by one with a chance of 50\% to the set of selected UTXOs. If the value of the selected UTXOs matches the target, the algorithm ends and outputs the selected UTXOs. However, if the value exceeds the target, the algorithm attempts to replace the last selected UTXO with a small UTXO to produce an even smaller set. In the second phase, the same operations are performed with the difference that every unselected UTXO is considered as an addition to the selected UTXO set. The procedure can be repeated iteratively to find better solutions. The algorithm is described in Algorithm~\ref{alg:knapsack}. The Knapsack algorithm has several advantages. In particular, it will decrease the size of the UTXO pool \ref{objective3} and increase privacy since only little information is revealed due to pseudo-randomly selected UTXOs \ref{objective2}. However, rapid reduction in the pool of UTXO could be an issue with respect to privacy \ref{objective2}. Another disadvantage is that it will not minimize the transaction fee as it uses a larger number of inputs \ref{objective1}.

\begin{algorithm}[t!]
\caption{Knapsack}\label{alg:knapsack}
\textbf{Input:} UTXO pool $U = \{u_1,...,u_n\}$ with $u_i^v \geq u_{i'}^v$ for $i<i'$\\
\textbf{Input:} Target $T>0$\\
\textbf{Output:} Selected UTXOs' set $S^\text{Knp}$
\vspace{-0.2em}
\begin{algorithmic}[1]
\Require $U^v > T$
    \State $Selected \leftarrow \{\}$
    \State $S^\text{Knp} \leftarrow \{\}$
    \State $bestVal \leftarrow \infty$
   % \State $rounds \leftarrow 10000$
    \State $targetReached \leftarrow false$
%\For{$i=1 \TO rounds$}
\For{$j=1 \TO 2$}
\If{$\neg targetReached$}
\For{$u\in U \setminus Selected$}
\State $randBool \leftarrow binaryRandom()$
\If{$(j=1 \AND randBool) \OR j=2$}
    \State $Selected$.\text{add}($u$)
    \If{${Selected}^v == T$}
        \State $targetReached \leftarrow true$
        \State $S^\text{Knp} \leftarrow Selected$
        \State \textbf{break}
    \EndIf
    \If{${Selected}^v > T$}
        \State $targetReached \leftarrow true$
        \If{${Selected}^v < bestVal$}
            \State $S^\text{Knp} \leftarrow Selected$
            \State $bestVal \leftarrow {S^\text{Knp}}^v$
            \State $Selected \leftarrow Selected \setminus \{u\}$
        \EndIf
    \EndIf
\EndIf
\EndFor
\EndIf
\EndFor
\State \textbf{return} $S^\text{Knp}$
%\EndFor
\end{algorithmic}
\end{algorithm}

\subsubsection{Branch and Bound}% - Bitcoin Core} 
The branch and bound (BnB) coin selection algorithm was proposed by Erhardt~\cite{Erhardt2016} and explained in  \cite{CoinSelectionForDummies2022PART2}. First, we need to introduce the concept of ``effective value''. As mentioned before, having more inputs and outputs in a transaction results in a higher transaction fee. For every selected input UTXO, there is a fee to be paid. We introduce the concept of effective value which is the original UTXO's value minus the transaction fee.  More precisely, suppose that there is a UTXO $u$ and the transaction fee per byte is $f$. Hence, the effective value of a UTXO is the UTXO's value $u^v$ minus the transaction fee per byte $f$ times the UTXO's size $u^s$. In other words, 
\begin{equation*}
    effVal(u) := u^v - fu^s. 
\end{equation*}
The advantage of considering effective values of UTXOs is that it keeps the target fixed throughout the process.

The BnB algorithm utilizes a depth-first search on a binary tree, where each node represents the inclusion or omission of a UTXO.  UTXOs are sorted in descending order of effective values, and the tree is explored deterministically, prioritizing the inclusion branch first.

Paths with a total effective value that exceeds the target are cut. Finding an exact match results in one less output and one less input for future transactions. Indeed, if an exact match is found, there is no change output in the current transaction, and this change output, in turn, does not need to be spent as input in future transactions. In total, we can save the cost of an input plus the cost of an output. We call this summation ``matchRange''. The algorithm is described in Algorithm~\ref{alg:bnb} and Algorithm~\ref{alg:bnb2}. If there is no match, there is a fallback logic that performs a random draw.

\begin{algorithm}[t!]
\caption{Branch and Bound --- $BnB(U, T, mc)$}\label{alg:bnb}
\textbf{Input:} UTXO pool $U = \{u_1,...,u_n\}$\\
\textbf{Input:} Target $T>0$\\
\textbf{Input:} Minimum change $mc$\\
\textbf{Input:} Rounds $rounds$\\
\textbf{Output:} Set of selected UTXOs $S^\text{BnB}$
\vspace{-0.2em}
\begin{algorithmic}[1]
\Require $\sum_i effVal(u_i)>T$
    \State $curentSelection \leftarrow \{\}$
    \State $rounds \leftarrow 1000$  
    \State $d \leftarrow 0$
    \State $curentSelection \leftarrow$  BnBRecursion($d$,$curentSelection$)
    \If{$currentSelection == \{\}$}
        \State $U' \leftarrow randomShuffle(U)$
        \While{${effValue(currentSelection)} < T + mc$}
        \For{$u' \in U'$}
        \State $currentSelection.\text{add}(u')$
        \EndFor
        \EndWhile
        \EndIf
    \State \textbf{return} $curentSelection$ as $S^\text{BnB}$
\end{algorithmic}
\end{algorithm}

\begin{algorithm}[t!]
\caption{Branch and Bound Recursion --- \\$BnBRecursion(d,currentSelection)$}\label{alg:bnb2}
\textbf{Input:} Depth level in search $d$\\
\textbf{Input:} Current UTXO Selection $currentSelection$\\
\textbf{Output:} New UTXOs Selection $currentSelection$
\vspace{-0.2em}
\begin{algorithmic}[1]
    \State $rounds \leftarrow rounds - 1$
    \State $targetForMatch \leftarrow$ T + Cost\_of\_Header + Cost\_per\_Output
    \State $matchRange \leftarrow$ Cost\_Per\_Input + Cost\_Per\_Output
    \State $utxoSorted \leftarrow sortDescending(U \setminus currentSelection)$  
    \State $currentEffValue=\sum_{i:u_i \in currentSelection} effVal(u_i)$
    \If{$currentEffValue > targetForMatch + matchRange$}
        \State \textbf{return} $\{\}$\\
    \textbf{else if} $currentEffValue \geq targetForMatch$ 
        \State \textbf{return} $currentSelection$\\
    \textbf{else if} $rounds \leq 0$ 
        \State \textbf{return} $\{\}$\\
    \textbf{else if} $d \geq |utxoSorted|$ 
       \State \textbf{return} $\{\}$
    \Else
        \If{$binaryRandom()==true$}
            \State $withThis \leftarrow BnBRecursion(d+1, currentSelection \cup \{utxoSorted[d]\})$
            \If{$withThis \neq \{\}$}
                \State \textbf{return} $withThis$ 
                \Else 
                \State $withoutThis \leftarrow BnBRecursion(d+1,currentSelection)$
                    \If{$withoutThis \neq \{\}$}
                        \State \textbf{return} $withoutThis$
                    \EndIf
            \EndIf
        \Else
        \State \textit{// As above but explore omission branch first} 
        \EndIf
    \EndIf
    \State \textbf{return} $curentSelection$
\end{algorithmic}
\end{algorithm}
\subsubsection{A Summary of Basic Algorithms}
We summarize the basic methods in Table~\ref{tab: basic algorithms} and show the objectives they achieve and to what extent. Basic algorithms are being used in different existing UTXO-based blockchain technologies such as Bitcoin and Cardano. However, as blockchain technology matures and applications proliferate, the shortcomings of these early algorithms become evident. Consequently, increasing emphasis has been placed on conducting thorough research and developing more advanced coin selection algorithms that address critical issues. This shift towards more sophisticated algorithms has played a crucial role in enhancing the overall performance and functionality of blockchain systems to satisfy all the mentioned objectives.
\begin{table}[t!]
\centering
\caption{Basic algorithms.}
\label{tab: basic algorithms}
\begin{tabular}{|l|c|c|c|c|c|}
\hline
\multirow{2}{*}{\textbf{Basic Algorithms}} & \multicolumn{5}{|c|}{\textbf{Objectives}} \\
& \ref{objective1} & \ref{objective2} & \ref{objective3}  & \ref{objective5} & \ref{objective6} \\
\hline
Greedy & $\circlerighthalfblack$  & $\mdwhtcircle$  & $\mdwhtcircle$  & $ \mdwhtcircle$  & $ \mdwhtcircle$  \\
Random Draw & $\mdwhtcircle$  & $\mdlgblkcircle$  & $\mdwhtcircle$  & $ \mdwhtcircle$  & $ \mdlgblkcircle$  \\  
Random-Improve & $\mdwhtcircle$  & $\circlerighthalfblack$  & $\mdlgblkcircle$  & $ \mdwhtcircle$  & $ \circlerighthalfblack$  \\  
Knapsack & $\circlerighthalfblack$  & $\circlerighthalfblack$  & $\mdlgblkcircle$  &  $ \mdwhtcircle$  & $ \mdwhtcircle$  \\  
Branch and Bound & $\circlerighthalfblack$  & $\mdlgblkcircle$  & $\mdwhtcircle$  &  $ \mdwhtcircle$  & $\mdwhtcircle$\\
\hline
\end{tabular}
\end{table}

\subsection{Advanced Algorithms}
We start this section by introducing basic terminologies. Let $\mathcal{T}=\{T_1,...,T_t\}$ \footnote{In the previous sections $\mathcal{T} = \{T\}$ was a singleton containing only one target $T$.} denote a set of payment requests. Furthermore, let $O = \{o_1,...,o_m\} \subseteq \mathcal{T}$ denote the set of outputs and for any $j \in [m]$, $o_j^\text{v}$ is the value of $o_j$ and $o_j^\text{s}$ the size of $o_j$. There is a change output $c$ with value $c^\text{v}$ and size $c^\text{s}$. We now define a transaction as a 3-tuple $(S^\text{alg}, O, c)$. Furthermore, let $D>0$ denote the dust threshold.

\subsubsection{Optimization Algorithm}
This section introduces an approach introduced primarily in \cite{Nguyen2018}. The optimization algorithm consists of two phases. The first phase aims to optimize the transaction size. The second phase focuses on minimizing the pool size. Before formulating the optimization problems, we start by introducing some notation. 

For any $i \in [n]$, the binary variable $x_i$ is 1 if $u_i$ is chosen as input and 0 otherwise. The transaction fee will be a product of a fixed fee rate $\alpha \geq 0$ and the transaction size. Let $\epsilon$ denote the minimum change output that is set to avoid creating a very small output.

The transaction size $y$ can be calculated as follows:
\begin{equation}
     y =  {\sum_{i\in [n]}{u_i^s x_i}+\sum_{j \in [m]}{o_j^\text{s}} + c^\text{s}}
\end{equation}
Note that $\sum_{i\in [n]}{u_i^\text{s} x_i} = |S^\text{Opt}$| which is the size of the UTXOs set  $S^\text{Opt}$ selected by the optimization method. Our objective is to minimize $y$ subject to the following constraints. In particular, the transaction size should be less than the maximum transaction size denoted by $M>0$. Also, each UTXO must have a sufficient value for consumption and all UTXOs outputs must be larger than the dust threshold $D>0$. Furthermore, the change size $c^\text{s} = \beta$ if $c^\text{v}>\epsilon$ and 0 otherwise. To minimize the transaction size in the first phase,  the optimization problem is as follows:
\begin{equation}
    \begin{aligned}
         \underset{x_i,y,m}{\min}  \quad & y \\
       \textrm{s.t.} \quad & y \leq M\\
        & \sum_{i \in [n]}{u_i^v x_i}= \sum_{j \in [m]}{o_j^v}+\alpha y + c^v  \\
        & \sum_{j \in [m]} o_j^v \geq D \\
        & c^s \leq \lfloor \frac{c^v}{\epsilon} \rfloor  \beta\\
        & x_i\in\{0,1\} \text{ for all } i \in [n]
    \end{aligned}
\end{equation}
We denote $y^\text{opt}$ as a solution to the above optimization problem which is the minimal transaction size. The second phase aims to minimize the size of the UTXO pool, that is, maximize the number of inputs in the transaction. The optimization problem has another constraint in addition to the constraints of the first phase. 
Let $\gamma \in (0,1)$. The optimization problem is as follows: 
\begin{equation}
    \begin{aligned}
         \underset{x_i,y,m}{\max}  \quad  &{\sum_{i \in [n]}{x_i}-\frac{c^s}{\beta}}\\
       \textrm{s.t.} \quad & y \leq M\\
        & \sum_{i \in [n]}{u_i^v x_i}= \sum_{j \in [m]}{o_j^v}+\alpha y + c^v  \\
        & \sum_{j \in [m]} o_j^v \geq D \\
        & c^s \leq \lfloor \frac{c^v}{\epsilon} \rfloor \beta \\
        & x_i\in\{0,1\} \text{ for all } i \in [n] \\
       \quad & y \leq (1+\gamma) y^\text{opt}
    \end{aligned}
\end{equation}
Note that if $\gamma$ is close to 0, we would like to keep the minimum transaction size obtained from the previous algorithm. If $\gamma$ is close to 1, a transaction of appropriate size is created by a number of UTXOs as large as possible. 

\subsubsection{Knapsack with Leverage}
The problem of Knapsack with leverage was studied by Diroff~\cite{Diroff2019}. The concept behind the standard Knapsack algorithm mentioned above involves finding a cost-effective and efficient transaction to handle a specified collection of pay requests. However, attempting to find a feasible solution to the Knapsack problem may not always succeed due to various reasons, such as the absence of a solution or the algorithm's inability to produce one within a designated time frame. It aims to address the entire problem by first attempting to find a solution using the Knapsack algorithm. If this attempt fails, it then resorts to utilizing the fallback solution.

\par The leverage solution, on the other hand, does not immediately rely on the fallback solution, but instead seeks to take advantage of the fact that when the standard Knapsack algorithm fails, there will be a change in the output of the transaction. The leverage solution strives to construct a transaction in such a way that the change output becomes a useful future UTXO. Essentially, it tries to create two transactions: one for processing the current pay requests and the other for handling a different set of pay requests. The goal is to ensure that the change output from the first transaction fits precisely into the second transaction, resulting in a change-free process. The paper studies the basic problem of two transactions and then extends it. In this work, we focus on the basic problem studied in \cite{Diroff2019}. 

\par We expand the definition of a transaction by adding a tip denoted by $r\geq 0$ that will be paid to the block producers as an incentive to collect the transaction from the pool for the block generation process. We assume that $\mathcal{T}$ is ordered by decreasing values. In particular, a transaction is a 4-tuple $(S, O, c, r)$. The size of the transaction is as follows:
\begin{equation*}
     x + y |S| + z |O| + z (1-\delta_{c,0}),
\end{equation*}
where $(x,y,z) = (10,148,34)$ are constants denoting the number of bytes required for metadata\footnote{Metadata is data that is describing other data. It could contain information about, e.g., an NFT or a transaction}, to record each input and to record each output \cite{Diroff2019}. $\delta_{c,0}$ is the Kronecker delta. We now start with the standard Knapsack problem where we denote $W (S^\text{Knp}, O, c, r)$ as the size of transaction generated by the algorithm.

\begin{equation}
    \begin{aligned}
         \min  \quad  &{W(S^\text{Knp}, O, c=0, r) \alpha + r}\\
       \textrm{s.t.} \quad & (S^\text{Knp}, O, c=0, r) \hspace{3 mm}\text{is a good and valid transaction}, 
    \end{aligned}
\end{equation}
where $\alpha$ is the market rate for the transaction fee per byte. Transaction $(S^\text{Knp}, O, c, r)$ is a valid transaction if 
\begin{center}
    $
     \begin{cases}
         \sum_{u\in S^\text{Knp}} {u} \geq \sum_{o\in O} {o} + W(S^\text{Knp}, O, c, r) \alpha\\
         \sum_{u\in S^\text{Knp}} {u}= \sum_{o\in O} {o} +W(S^\text{Knp}, O, c=0, r) \alpha+c+r
     \end{cases}
     $
\end{center}
and is a good transaction if
\begin{center}
    $
     \begin{cases}
          c=0\hspace{3mm} \text{and } 0\leq r \leq H\\
          \text{or}\\
         c\geq D \hspace{3mm} \text{and} \hspace{3mm} r=0
     \end{cases}
     $
\end{center}
where $H$ is the maximal overpayment amount.

The attempt at finding any feasible solution to the Knapsack problem may fail for several reasons (e.g. there is no solution or the algorithm fails to produce one in a certain allotted time period). Once the standard knapsack algorithm fails, it is known that there will be a change output in the transaction. In this case, the leverage solution attempts to construct two transactions, one processing the current pay requests $O_1$ and the other one processing some other set $O_2$, so that the change output of the first transaction fits precisely in the second, making it change-free. This method is called the Knapsack with leverage and is defined in (\ref{npskwithlev}). Let $\tau_1: = (S^\text{KnpLv}_1, O_1, c_1, r_1)$ and $\tau_2:= (S^\text{KnpLv}_2 \cup \{c_1\}, O_2 \subseteq \mathcal{T}\setminus O_1, c_2=0, r_2)$ be two transactions.

      \begin{equation}\label{npskwithlev}
          \begin{aligned}
         \min  \quad  &({\tau_1^\text{s} \alpha + r_1, \tau_2^\text{s} \alpha + r_2})\\
       \textrm{s.t.} \quad & \tau_1 \hspace{3 mm}\text{is a good and valid transaction} \\
        &  \tau_2 \hspace{3 mm}\text{ is a good, change-free and  valid transaction.}
     \end{aligned}
     \end{equation}
The advantage of the Knapsack with leverage algorithm is that the transaction fee is minimized \ref{objective1}. It looks at several transactions and hence considers the long run, and also considers confirmation time. However, it does not focus on privacy \ref{objective2}.

\subsubsection{Myopic and strategic optimization}
The myopic and strategic optimization was studied in \cite{Abramova2020}. Given the initial pool of UTXO $U$, a subset $S^\text{alg}_1 \subseteq U$ of UTXO is selected to meet the target $T_1$ in the first transaction in the first period. In the second period, there is a target $T_2$ that must be reached by choosing a subset $S^\text{alg}_2 \subseteq (U \setminus S^\text{alg}_1) \cup \{c_1\}$ where $c_1$ is the change output of the first transaction. The article considers two settings based on the information about $T_2$ as follows. 
\begin{enumerate}
    \item Myopic optimization: There is no information about $T_2$ in the first period.
    \item Strategic optimization: $T_2$ is known in the first period.
\end{enumerate}
The myopic optimization problem is as follows:
      \begin{equation}
          \begin{aligned}
          \underset{x_i}{\min} \quad &|S^\text{Myop}_i|  \\
        \quad \textrm{s.t.}  \quad & T_i \leq \sum_{i\in [n]} u_i^\text{v} x_i \text{ for } i=1,2\\
       & x_i\in\{0,1\} \text{ for all } i \in [n]
     \end{aligned}
     \end{equation}
For the strategic setting, we introduce $\lambda\in [0,1]$ as the preference for privacy over lower transaction fees and $A_i$ as the set of all UTXO addresses in the $i$ th transaction ($i=1,2$). Furthermore, the strategic optimization problem is as follows: 
      \begin{equation}
          \begin{aligned}
         & \min \quad  (1-\lambda) |S^\text{Strat}_1 \cup S^\text{Strat}_2| + \lambda \mathbbm{1}_{(A_1 \cap A_2 \neq \emptyset \vee c_1 \in S^\text{Strat}_2 )} \\
       & \quad \textrm{s.t.}  \quad T_i \leq ({S^\text{Strat}_i})^v \text{ for } i=1,2,
     \end{aligned}
     \end{equation}
where $\emptyset$ refers to the empty set and $\mathbbm{1}_{x} =\begin{cases}
    1 , \text{ if $x$ holds}\\
    0, \text{ otherwise}.
\end{cases}$
The myopic approach focuses on minimizing the transaction size, and hence the transaction fee, whereas the strategic approach additionally focuses on improving privacy.

\subsubsection{Greedy and genetic algorithm}
\begin{algorithm}[t!]
\caption{Greedy and Genetic --- $GrGe(U)$}\label{alg:greedy_genetic}
\textbf{Input:} UTXO pool $U = \{u_1,...,u_n\}$ with $u_i^v \geq u_{i'}^v$ for $i<i'$\\
\textbf{Input:} Target $T>0$\\
\textbf{Input:} Number of rounds $K>0$\\
\textbf{Input:} Size $M>0$\\
\textbf{Output:} Set of selected UTXOs $S^\text{GrGe}$
\vspace{-0.2em}
\begin{algorithmic}[1]
\Require $U^v \geq T$
\If{$U^v == T$}
    \State \textbf{return} $U$
\EndIf
\If{$U^v > T$ \AND $\max(U) > T$}
    \State \textbf{return} $\min(\{u \in U | u^v \geq T\})$
\EndIf
\State $\mathcal{I} \leftarrow \{\}$
\State $best \leftarrow G(U)$\footnotemark
\State $\mathcal{I}.\text{add}(best)$
\For{$i=1 \TO M-1$}
    \State $I \leftarrow random(\mathcal{P}(U))$
    \State $\mathcal{I}.\text{add}(I)$
    \If{$f(I) > f(best)$}
        \State $best \leftarrow I$
    \EndIf
\EndFor
\If{$best^v == T$}
    \State $S^\text{GrGe} \leftarrow best$
    \State \textbf{return} $S^\text{GrGe}$
\EndIf
\For{$k = 1 \TO K-1$}
    \State Generate new set $\mathcal{I}$ using randomness
    \For{$I \in \mathcal{I}$}
        \If{$f(I) > f(best)$}
            \State $best \leftarrow I$
        \EndIf
    \EndFor
    \If{$best^v == T$}
        \State $S^\text{GrGe} \leftarrow best$
        \State \textbf{return} $S^\text{GrGe}$
    \EndIf
\EndFor
\State \textbf{return} $S^\text{GrGe}$
\end{algorithmic}
\end{algorithm}\footnotetext{Here we use the greedy algorithm $G(U)$ described in Algorithm~\ref{alg:greedy}.}

Originally, the genetic algorithm was introduced in \cite{Holland1975} and follows the evolution of a population to solve the optimization problem using an exhaustive search. The algorithm was recently studied in the context of coin selection \cite{Wei2023}. The algorithm in \cite{Wei2023} is as follows. First, the least number of UTXOs as transaction inputs is determined using the greedy algorithm. This number is then used in the objective function of the genetic algorithm. The optimal solution is then searched using the genetic algorithm. We summarize the procedure in the following way. For each $u_i^v\geq u_{i'}^v$ for $i,i' \in [n]$ with $i<i'$ where $u_i, u_{i'} \in U$. 

Furthermore, the function $random(\mathcal{P}(U))$ randomly selects a subset of $U$ which is the summation of their values that exceed the target. Possible solutions are compared using a ``fitness'' function, which is defined as follows. Let $B \in \mathcal{P}(U)$, then the fitness of $B$ is given by
\begin{equation*}
    f(B) = \frac{1}{B^v - T + |B|}.
\end{equation*}
The greedy and genetic algorithm works as follows. Let $M,K > 0$ be given. First, it is checked whether there is enough balance in the UTXO pool and whether the balance matches the target or not. If the balance does not match the target, the algorithm performs the greedy algorithm to get an initial best solution, that is, a subset of $U$ which has a balance greater than or equal to the target value. The algorithm then randomly draws $M-1$ additional subsets of $U$ such that the sum of their values exceeds the target value. In total, we now have $M$ possible solutions. For each of the $M-1$ randomly drawn subsets, we calculate the fitness and compare it to the fitness of the best initial solution. Whenever there is a subset with better fitness, that subset will be the best. For the next $K-1$ iterations, the initial $M$ subsets are changed by randomly adding and removing the UTXOs, and again the fitness is calculated and compared to the best. Once there is a subset with a value matching the target value, we stop or otherwise, after all iterations, the best solution is returned. The algorithm is described in Algorithm~\ref{alg:greedy_genetic}. 

Note that when generating a new set $\mathcal{I}$, the randomness is created using three procedures. First, $I\in\mathcal{I}$ remains in the next round with probability proportionally to fitness. Second, from every pair $I,I'\in \mathcal{I}$ generate a new pair using randomness (called a single-point crossover). And lastly, random invert digits in the binary string of each element with some probability (called mutation).

\par The advantage of this algorithm is that it minimizes transaction fees \ref{objective1} as it reduces the number of input UTXOs. However, it does not minimize the size of the UTXO pool.

\subsubsection{A Summary of Advanced Algorithms}
We summarize the advanced algorithms in Table~\ref{tab: advanced algorithms} and describe the objectives they achieve and to what extent. 

\begin{table}[t!]
\centering
\caption{Advanced algorithms.}
\label{tab: advanced algorithms}
\begin{tabular}{|l|c|c|c|c|c|}
\hline
\multirow{2}{*}{\textbf{Advanced Algorithms}} & \multicolumn{5}{|c|}{\textbf{Objectives}} \\
& \ref{objective1} & \ref{objective2} & \ref{objective3} & \ref{objective5} & \ref{objective6} \\
\hline
Optimization & $\mdlgblkcircle$  & $\mdwhtcircle$  & $\mdlgblkcircle$    & $ \mdwhtcircle$  & $ \mdwhtcircle$  \\  
Knapsack w/ Lev.& $\mdlgblkcircle$  & $\mdwhtcircle$  & $\mdwhtcircle$   & $ \circlerighthalfblack$  & $ \mdwhtcircle$  \\  
Myopic and Str. Opt. & $\mdlgblkcircle$  & $\circlerighthalfblack$  & $\mdwhtcircle$   & $ \mdwhtcircle$  & $ \mdwhtcircle$  \\  
Greedy and Genetic & $\mdlgblkcircle$  & $\mdwhtcircle$  & $\mdwhtcircle$  & $ \mdwhtcircle$  & $\mdwhtcircle$\\
\hline
\end{tabular}
\end{table}

\section{Performance Evaluation} \label{sec: perfomance}

\par In this section, we first summarize our findings in Table~\ref{tab: all} based on the previous discussions in Table~\ref{tab: primitive algorithms}, Table~\ref{tab: basic algorithms}, and Table~\ref{tab: advanced algorithms}. Then, we review the performance of those algorithms that are already deployed in existing blockchains. 

\begin{table}[t!]
\centering
\caption{Summary of all coin selection algorithms.}
\label{tab: all}
\begin{tabular}{|l|c|c|c|c|c|}
\hline
\multirow{2}{*}{\textbf{Algorithm}} & \multicolumn{5}{|c|}{\textbf{Objectives}} \\
& \ref{objective1} & \ref{objective2} & \ref{objective3} & \ref{objective5} & \ref{objective6} \\
\hline
\hline
First-In-First-Out (FIFO) & $\mdwhtcircle$  & $\circlerighthalfblack$  & $\mdwhtcircle$  &  $ \mdwhtcircle$  & $ \mdwhtcircle$  \\  
Last-In-First-Out (LIFO) & $\mdwhtcircle$  & $\circlerighthalfblack$  & $\mdwhtcircle$  &  $ \mdwhtcircle$  & $ \mdwhtcircle$  \\  
Highest Value First (HVF) & $\mdlgblkcircle$  & $\circlerighthalfblack$  & $\mdwhtcircle$  & $ \mdwhtcircle$  & $ \mdwhtcircle$  \\  
Lowest Value First (LVF) & $\mdwhtcircle$  & $\mdwhtcircle$  & $\mdlgblkcircle$  & $ \mdwhtcircle$  & $\mdwhtcircle$  \\ 
Highest Priority First (HPF) & $\mdlgblkcircle$  & $\mdlgblkcircle$  & $\mdwhtcircle$  &  $ \mdwhtcircle$  & $ \mdwhtcircle$  \\
\hline
\hline
Greedy & $\circlerighthalfblack$  & $\mdwhtcircle$  & $\mdwhtcircle$  & $ \mdwhtcircle$  & $ \mdwhtcircle$  \\
Random Draw & $\mdwhtcircle$  & $\mdlgblkcircle$  & $\mdwhtcircle$  & $ \mdwhtcircle$  & $ \mdlgblkcircle$  \\  
Random Improve & $\mdwhtcircle$  & $\circlerighthalfblack$  & $\mdlgblkcircle$  & $ \mdwhtcircle$  & $ \circlerighthalfblack$  \\  
Knapsack & $\circlerighthalfblack$  & $\circlerighthalfblack$  & $\mdlgblkcircle$  &  $ \mdwhtcircle$  & $ \mdwhtcircle$  \\  
Branch and Bound & $\circlerighthalfblack$  & $\mdlgblkcircle$  & $\mdwhtcircle$  &  $ \mdwhtcircle$  & $\mdwhtcircle$\\
\hline
\hline
Optimization & $\mdlgblkcircle$  & $\mdwhtcircle$  & $\mdlgblkcircle$    & $ \mdwhtcircle$  & $ \mdwhtcircle$  \\  
Knapsack with Leverage & $\mdlgblkcircle$  & $\mdwhtcircle$  & $\mdwhtcircle$   & $ \circlerighthalfblack$  & $ \mdwhtcircle$  \\  
Myopic and Str. Opt. & $\mdlgblkcircle$  & $\circlerighthalfblack$  & $\mdwhtcircle$   & $ \mdwhtcircle$  & $ \mdwhtcircle$  \\  
Greedy and Genetic & $\mdlgblkcircle$  & $\mdwhtcircle$  & $\mdwhtcircle$  & $ \mdwhtcircle$  & $\mdwhtcircle$\\
\hline
\end{tabular}
\end{table}

\par To model the approximate real-world behavior of a wallet, we need to incorporate deposits and payments into our environment. The deposit value, which we simply refer to as the deposit, is added to the wallet's UTXO balance. The payment value, which we refer to as the target, is subtracted from the wallet's UTXO balance. Wallets usually involve a number of incoming and outgoing transactions. This is taken care of by incorporating deposits and targets.

The environment is set up as follows. We model the case of a wallet with an initial UTXO value of 100,000 tokens. We run 10,000 iterations for each of the different algorithms and compare their performance. In each iteration, a random target is drawn, and the input UTXOs are chosen according to the underlying wallets' algorithm. The change UTXO is added to the wallet's UTXO pool. Additionally, in each iteration, three deposits are drawn and added to the UTXO pool. The targets and deposits are the same for all algorithms, and the deposit/target ratio is 3:1 \cite{Edsko2018, Erhardt2016}. Note that when changing the ratio, we need to make sure that the average of the deposits and the targets are still balanced. Otherwise,  the average of deposits or targets will increase in the long run, which is not desirable. We rather want to keep the balance almost the same when investigating different coin selection algorithms.

\par We study the system considering two different distributions for deposits and targets. First, the deposits and targets are drawn from a Normal distribution \cite{Edsko2018,Erhardt2016}. Specifically, the deposits are drawn from a Normal distribution with a mean of $1,000$ and a standard deviation of $250$. Also, the targets are drawn from a Normal distribution with a mean of $3,000$ and a standard deviation of $500$. Note that the averages of deposits and targets are equal and keep the wallets' balances around the initial balance. Second, we consider the case of a memoryless distribution. In particular, the deposits and targets are drawn from a Poisson distribution with mean $1,000$ and $3,000$, respectively.

\begin{figure}[tbp]
    \centering
    \includegraphics[width = 0.45\textwidth]{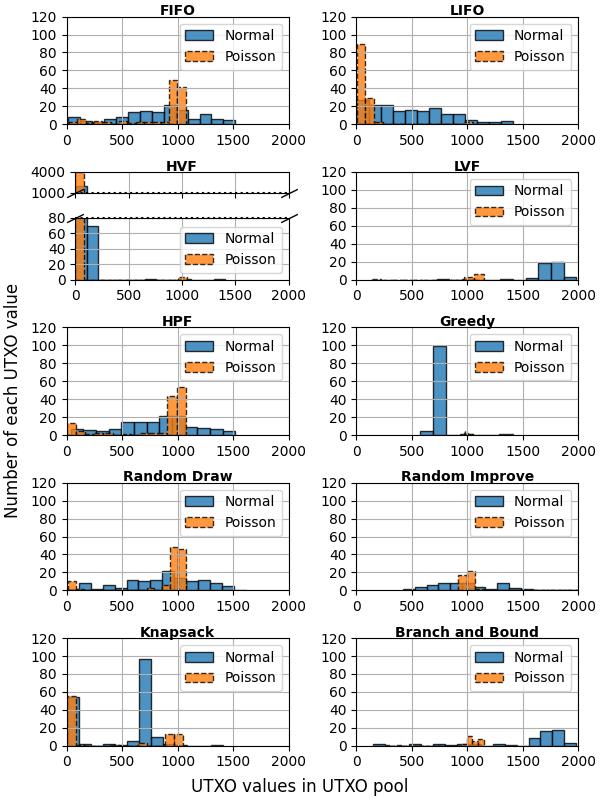}
    \caption{The number of each UTXOs value in the UTXO pools for primitive and basic coin selection algorithms.}
\label{fig:PoolValues_both}
\end{figure}

\par \par All algorithms start with a single initial UTXO of a value of 100,000 tokens. Over time, the wallets accumulate a number of UTXOs of different values which come from the deposits and from the change output after each iteration. Fig.~\ref{fig:PoolValues_both} shows the number of each UTXO value in the pool. According to the objective \ref{objective6}, it is desired to have a wide range of distributed UTXO values. Additionally, we would like UTXO values that are not too small (i.e., dust).
The range of values in Fig.~\ref{fig:PoolValues_both} is depicted to be up to 2,000. However, there are single UTXOs of larger value for some algorithms. In particular, for the Normal (Poisson) distribution, LIFO, LVF, and Greedy have each one UTXO with a value of around 30,000 (90,000). Knapsack and Branch\&Bound each have a UTXO with a value of around 20,000 (70,000), and the random improve algorithm has several UTXOs with a value of around 5,000 (4,000). Hence, qualitatively we have the same trends for both Normal and Poisson distributions. However, the interesting question is how many small UTXOs (dust) end up in a wallet. We clearly see that the HVF accumulates the largest amount of dust. In fact, while for the Normal distribution, the number of small UTXOs is around 2,000, the number is twice as much under the Poisson distribution. The Greedy and Knapsack algorithms also have many UTXOs of approximately the same size, instead of more distributed values, as is the case for the other algorithms.

\par
\begin{figure}
\centering
    \includegraphics[width =0.45\textwidth]{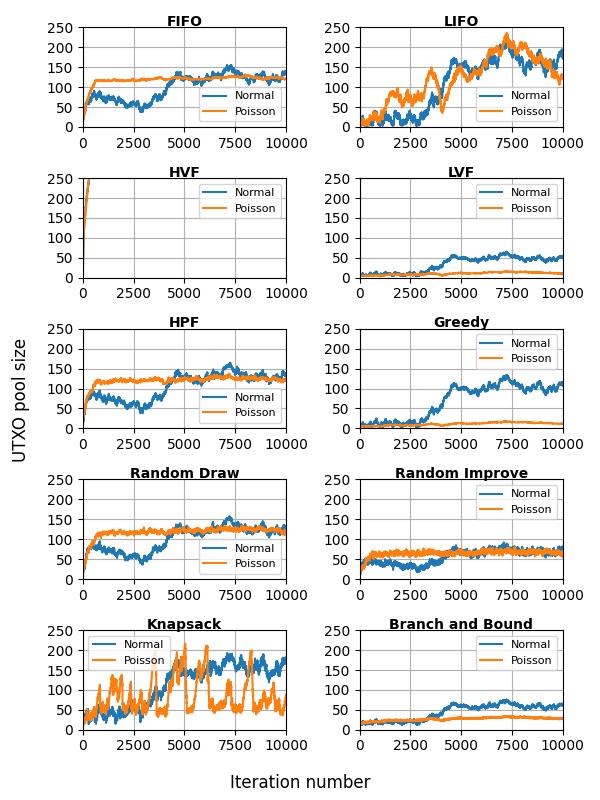}
    \caption{Size of the UTXO pools in different iterations for primitive and basic coin selection algorithms.}
    \label{fig:PoolSize_both}
\end{figure}

\begin{figure}[t!]
    \centering
    \includegraphics[width = 0.45\textwidth]{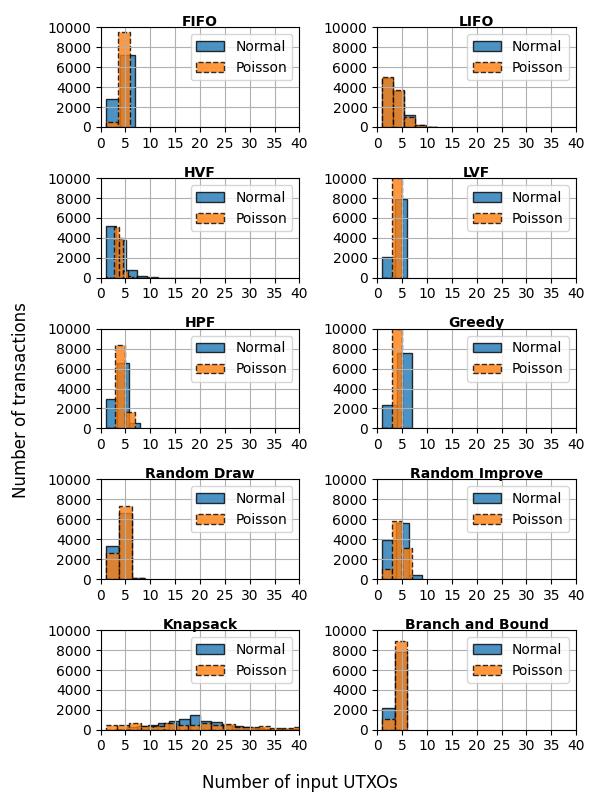}
    \caption{Number of transactions versus the number of input UTXOs for primitive and basic coin selection algorithms.}
    \label{fig:InputUTXOs_both}
\end{figure}

 The number of UTXOs in a wallet is depicted in Fig.~\ref{fig:PoolSize_both}. According to the objective \ref{objective3}, it is desired to have a low number of UTXOs in the pool. We clearly observe that all algorithms except HVF follow approximately the same trend. Under the Poisson distribution, the pool size remains very low for both Greedy and LVF, whereas for Normal distribution, it is more than double in size. In fact, under the Poisson distribution, both algorithms keep the pool size constantly low and lower than all other algorithms. For HVF, the number of UTXOs in the UTXO pool explodes compared to the other algorithms for both distributions. For Knapsack we observe that under Poisson distributed targets and deposits, the pool size fluctuates between 50 and 200 UTXOs. This is due to the fact that the number of input UTXOs is widely distributed and close to 150. That is, for several transactions, approximately 150 UTXOs are used as the input and therefore dramatically shrink the pool size. The distribution of the number of input UTXOs is depicted in Fig.~\ref{fig:InputUTXOs_both}. However, we represent it only up to 40 inputs to better compare it with other algorithms.

\par
\begin{figure}
\centering    
    \includegraphics[width = 0.45\textwidth]{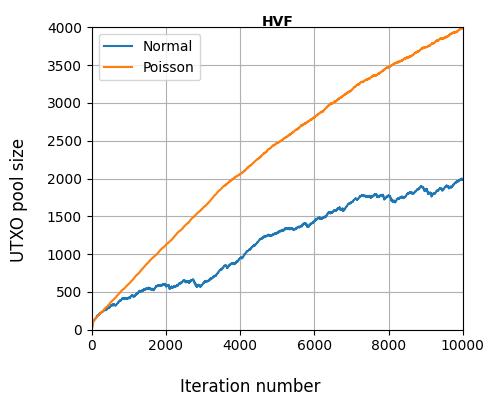}
    \caption{Size of the UTXO pool in different iteration for the HVF algorithm.}
    \label{fig:PoolSizeHVF_both}
\end{figure}

 Fig.~\ref{fig:PoolSizeHVF_both} shows the evolution of the pool size and we observe that the pool size under HVF is more than 10 (20) times larger than the pool sizes of the other algorithm under Normal (Poisson) distribution. Using HVF, the majority of UTXOs after 10,000 iterations are dust UTXOs, i.e. UTXOs with a tiny value. All algorithms, other than HVF, seem to stabilize the number of UTXOs in their pools.

Fig.~\ref{fig:InputUTXOs_both} describes the number of input UTXOs needed to pay the target in each iteration. According to the objective \ref{objective1}, the transaction fee should be minimized, which is the same as minimizing the number of inputs. In terms of privacy, a wider distribution of the number of inputs is desired. We observe that all algorithms except Knapsack use a small number of inputs, whereas Knapsack is widely distributed. In fact, Knapsack also has transactions with more than 100 inputs. This trend is observed under both distributions.

In general, we can conclude that the choice of a coin selection algorithm requires a deep understanding of the desired objectives. None of the coin-selection algorithms studied meets all objectives. Therefore, it requires a careful choice based on the underlying system and its requirements. 
For example, if privacy is a desired objective, then Knapsack would be the best choice, as the number of inputs is widely distributed. However, Knapsack leads to higher transaction fees as the transaction size increases with the number of input UTXOs. These trade-offs are the basis of the choice of a coin selection algorithm. 

\section{Conclusion}\label{sec: conclusion}
In this paper, we studied various coin selection algorithms and defined their objectives, classifications, and performance characteristics. Initially, we provided a list of the desired objectives for coin selection algorithms, such as minimizing transaction fees and transaction size, improving privacy, minimizing pool size, and reducing confirmation time. We then reviewed the coin selection algorithms, classifying them into three distinct categories: primitive, basic, and advanced. The primitive category refers to algorithms discussed and conceptualized during the early stages of the blockchain era. The basic category represents a significant advancement in terms of usability and effectiveness, catering to the growing demands and complexities of the blockchain domain. The advanced category pivots to more sophisticated algorithms, playing a crucial role in increasing the overall performance and functionality of blockchain systems. However, these advanced mechanisms have not yet seen widespread adoption in practical software systems. Finally, we compared the performance of the algorithms and discussed the advantages and disadvantages of each, providing a comprehensive view of coin selection algorithms within the blockchain space. This comparison not only highlights the evolutionary trajectory of these algorithms, but also serves as a guide for selecting the appropriate algorithm based on specific requirements and constraints. The final conclusion is that, at this moment, there is no coin selection algorithm that meets all the required objectives. Consequently, there is a significant gap in this area that future research needs to address.
\bibliographystyle{ieeetr}
\bibliography{references}
\end{document}